# Securing the Data in Clouds with Hyperelliptic Curve Cryptography

Debajyoti Mukhopadhyay
Department of Information Technology
Maharashtra Institute of Technology
Pune, India
debajyoti.mukhopadhyay@gmail.com

Ashay Shirwadkar
Department of Information Technology
Maharashtra Institute of Technology
Pune, India
ashayshirwadkar12@gmail.com

Pratik Gaikar
Department of Information Technology
Maharashtra Institute of Technology
Pune, India
pratik.gaikar2008@gmail.com

Tanmay Agrawal
Department of Information Technology
Maharashtra Institute of Technology
Pune, India
tanmay0208@gmail.com

*Abstract*—**In today's world, Cloud computing has attracted research communities as it provides services in reduced cost due to virtualizing all the necessary resources. Even modern business architecture depends upon Cloud computing .As it is a internet based utility, which provides various services over a network, it is prone to network based attacks. Hence security in clouds is the most important in case of cloud computing. Cloud Security concerns the customer to fully rely on storing data on clouds. That is why Cloud security has attracted attention of the research community. This paper will discuss securing the data in clouds by implementing key agreement, encryption and signature verification/generation with hyperelliptic curve cryptography**.

## I. INTRODUCTION

Cloud computing is becoming one the important topic in industry and academia with fast development of storage, network and software. It is also considered to be a promising internet based computing platform. The National Institute of Standards and Technology (NIST) define cloud computing as "a model for user convenience, on-demand network access contributes the computing resources (e.g. network, storage, application, servers and services) that can be rapidly implemented with minimal management effort or service provider interference" [2].

The users can access the cloud data and application at anytime and anywhere. The cloud contains large number of servers required to deliver scalable and reliable on-demand services. Many companies provide the cloud computing resource such as Cisco, Google, Microsoft, Marketo, Intacct, Amazon and VMware.The cloud has the elastic character and system allocation can get enlarge or shrink according to the requirement. The cloud also has the scalability and the cloud can scale higher for peak demand and lower for lighter demand.

The information in the cloud platform is very important for the client, the hacker pays more attention to hack that information because of this system must be protected more carefully than the traditional method.

Cloud computing security, performance and availability are most critical issue of the cloud computing resource and security being the most important [3].Using cloud computing can help in keeping one's IT budget to the minimum value. Emphasis on the cost and performance gain of cloud computing overshadow some of the fundamental security and privacy concerns.

## II. CLOUD SERVICES

Cloud computing service can be categorized into 4 main Services section: Infrastructure as a Service (IaaS), Platform as a Service (PaaS), Storage as a Service (StaaS) and Software as a Service (SaaS).

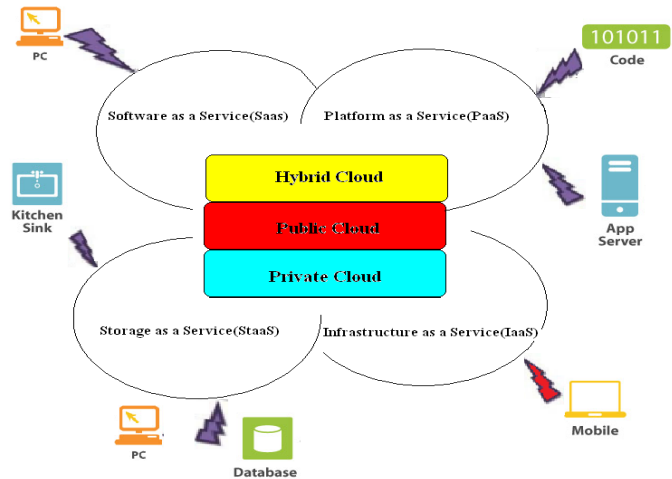

*Figure1: Cloud Computing Map*

## A. Infrastructure as a Service (IaaS)

It is the capability provided to the clients to use the Cloud computing vendor's system, networks and other basic computing resources where the client is able to deploy and run their system on it. The service vendor's owns the resources and is responsible for running, housing and maintaining it. The client only pays per-use basis. This greatly reduces the huge cost in initial investment of computing hardware such as servers, processing power and networking devices. Infrastructure as a Service is also referred to as Hardware as a Service (Haas).
E.g.: Amazon Web services, AT&T, Rackspace cloud, Terremark Enterprise Cloud and Windows Live Skydrive.

## B. Platform as a Service (PaaS)

It is the capability provided to the clients to use the cloud computing vendor's infrastructure to deploy and configure software or application on it. In this model, the client creates the software using tools and/or libraries from the vendor. The platforms offered include application design, development tools, deployment testing platform and configuration management.
E.g.: Google Application Engine, Microsoft Azure, Red hat OpenShift

## C. Storage as a Service (StaaS)

It is the capability provided to the clients to use the cloud computing vendor's storage and also enables cloud applications to use storage more than their limited servers. It rent storage space on a cost-per-gigabyte-stored or cost-per-data-transfer depending on the client requirements [5]. Storage as a service facilitate organization to reduce the storage amount while boosting the agility of storage framework.
E.g.: Amazon S3 and Nirvanix

## D. Software as a Service (SaaS)

It is the capability provided to the clients to use the Cloud computing vendor's dedicated applications running on Cloud infrastructure. The application has confine functionality and its core pack can be increased and decreased allowing of simple customization which is accordingly billed. SaaS is the most popular choice among users and it is also referred to as Software as a Product. SaaS sales have increased from $10 billion in 2010 to $12.1bn in 2011. It is estimated that SaaS revenue will reach $21.3bn by 2015[1].
E.g.: Online word processing and spreadsheet tools, Oracle, Microsoft office 365, LinkedIn.

## III. CLOUD DEPLOYMENT MODEL

Unconcerned of the service model utilized, there are three deployment models for cloud computing:

## A. Private Cloud

The cloud platform is solely operated for a specific organization. It may be handled and controlled by the same organization or a third party person. Private cloud provides the maximum enterprise control over deployment or use.

## B. Public Cloud

The public cloud is available to users to access and register the accessible platform via interface using mainstream web browser. Public cloud model can be free or provide on a pay-per-usage model. Public cloud is less secure than other cloud models

## C. Hybrid Cloud

The hybrid cloud platform is a composition of private & public cloud which is centrally managed and provisioned as a single unit. Hybrid cloud adds more secure control of data and application and allows benefits of both private as well as public clouds [5].

## IV. CLOUD SECURITY CHALLENGES

Institutions are rapidly moving onto cloud because of the availability of best resources in the market and also reduction in their working amount drastically. The cloud computing has no boundaries because of it more and more data is moved to the cloud increasing security concerns.

Many of the challenges should be addressed through management initiatives. Some of the important issues are mentioned below,

- **Data Loss and Leakage:** Data Loss and Leakage is one important security issue. Deletion and loss of data due to accident or disaster (flood, fire, and earthquake) will cause a number of the problems to the company whose information was stored. If the lost information is not retrieved, the company will have to face a huge financial loss. [7]

- **Denial of Service (DoS):** It is also one of major issue wherein the client is granted partial or no access to their information. DoS attacks are done by one person or system and leads to server overload.

- **Eavesdropping:** Connection eavesdropping means that an attacker can look your online activities and replay a particular transmission to get into your private data. It can also lead password cracking.

- **Malicious insiders:** It is a threat to company that comes from people within the company like employees and associates. This threat leads to fraud, theft of confidential or financial data. [7]

- **Lack of safety standards:** Till today, less attention is given defining mutual safety standards.

- **Compatibility** between different cloud platforms is also an issue. If a client wishes to move from one cloud to another the compatibility ensures that there is no issue of data transfer.[8]

## V. HYPERELLIPTIC CURVE IN CRYPTOGRAPHY

Hyperelliptic curves are class of algebraic curves that can be viewed as generalizations of elliptic curves. Hyperelliptic curve cryptography is defined over curves whose genus ≥ 1. The curve with genus 1 is commonly known as elliptic curve. ECC (Elliptic Curve Cryptography) is proven to be better pubic key cryptosystems than RSA , DSA. This is because decent amount of literature are published in that field, but same is not true for hyperelliptic curve , it has not received much attention by research communities. It has been restricted by academic purpose only.

Let k be a field. The general equation of hyperelliptic curve C of genus g over k is

C: y2 + h(x)y = f(x)

Where
- *h(x)* is a polynomial of degree ≤ *g* over **F**
- *f(x)* is a monic polynomial of degree *2g+1*
Over **F**
- certain further conditions

For example figure shows curve

C : $y^2 = x^5 - 5x^3 - 4x - 1$ over ℚ, genus g = 2

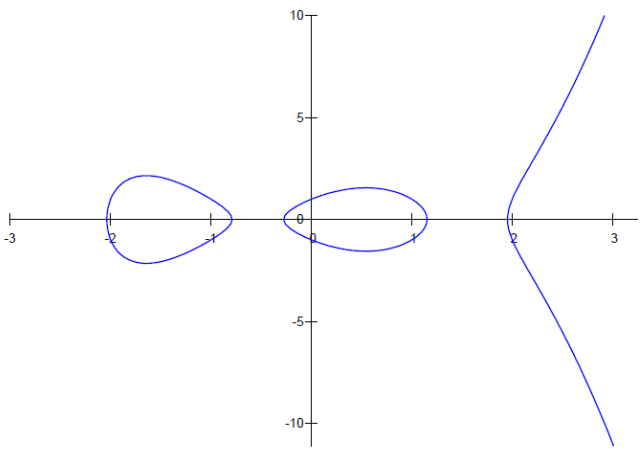

*Figure2: Hyperelliptic Curve*

Genus of a curve is a number of nonintersecting simple closed curves that can be drawn on the surface without separating it. It is equal to the number of handles on it. So for hyperelliptic curve genus is ≥ 1.

### A. Divisor

A divisor D is a formal sum of points P ∈C [17]:

$$D = \sum_{P \in C} m_p P$$

With $m_p \in$ Z and for all but finitely many $m_P = 0$.
The degree of D is the integer deg(D) = $\sum_{P \in C} m_p$.
The order of D at P is the integer $\text{ord}_P$ (D) = $m_P$.

The divisors form a group under addition. The group of divisors of Curve C is denoted by Div(C). We can add two divisors as follows [13].

$$\sum_{P \in C} m_p P + \sum_{P \in C} n_p P = \sum_{P \in C} (m_p + n_p) P$$

### B. Semi-Reduced Divisor

A semi-reduced divisor is a divisor of the form [17]:

$$D = \sum_i m_i P_i - \left(\sum_i m_i\right) \infty$$

Where each mi ≥ 0 and all the $P_i$'s are finite points such that if P ∈supp(D),then $\tilde{P}$ ∈supp(D) unless P is special in which case $m_i$ = 1[13].

### C. Reduced Divisor

Let $D = \sum_i m_i P_i - (\sum_i m_i)\infty$ be a semi reduced divisor. We call D to be a reduced divisor, if it satisfies the following property. Where g is genus of curve,

$$\sum m_i \leq g$$

There are two ways of Divisor Representation. Firstly Point Representation or Explicit Representation. this is the simplest form of representation. This is the representation which directly follows from the definition of the divisor word by word. Second is Mumford Representation, This is the representation which is mostly used for the computing purposes. It has advantages over normal representation.

Now as we have got familiar with addition of Divisors, lets apply this concept to Hyperelliptic curve. Consider the same curve equation,

C : $y^2 = x^5 - 5x^3 - 4x - 1$ over ℚ, D = [R1] + [R2] + [R3]

where R1 = (-2,-1) , R2 = (2,-1) , R3 = (3,-11).These points lie on the quadratic y = $-2x^2$ + 7 (shown by pink line).

This quadratic meets C in two other points P1, P2(green dots).Thus ,

$$[R1] + [R2] + [R3] + [P1] + [P2] = 0 \text{ in } Jac_C(\mathbb{Q})$$

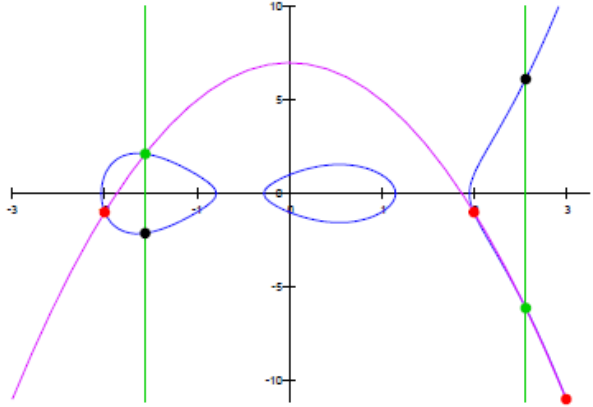

Figure3: Point addition in Hyperelliptic Curve

∴ [R1] + [R2] + [R3] = [B1] + [B2] with B1 = [$\overline{P1}$],B1 = [$\overline{P2}$] (black dots).

Hyperelliptic curve cryptography [HCC] is type of public-key cryptography which is successor of ECC. Every user has pair of public key and private key. Private key is used for decryption / signature generation whereas Public key is used for encryption/signature verification.

## VI. PROCEDURE FOR DATA SECURITY USING HCC

Consider we have two organizations A and B. Both act as public clouds. They have their own data and software.

Consider situation that organization A wants to send some data on demand of B. On request of B , A will search the data in its database and will retrieve the same. After that A will use encryption scheme discussed below and will sign the document using its private key. Thus encrypted message along with signature is sent by A. At B , firstly it will verify the sign using A's public key. If verified will decrypt the ciphertext and get the original message.

## VII. ALGORITHM FOR DATA SECURITY USING HCC

We propose three types of schemes based on hyperelliptic curve cryptography. They are key agreement, encryption and signature schemes.

### A. Key Agreement

The Diffie-Hellman key agreement was formulated for the multiplicative group of numbers modulo a prime, but it can easily be adjusted to general groups.

Let G be a group whose elements can be represented in an efficient way, and in which the group operations can be evaluated efficiently as well. The group is jacobians of hyperelliptic curves.[16]

First, we assume that there are the following publicly known system parameters:
– The group G.
– An element R ∈ G of large prime order r.

The steps that Organization A performs are the following:
1. Choose a random integer a ∈ [1, r − 1].
2. Compute P = aR in the group G, and send it to B.
3. Receive the element Q ∈G from B.
4. Compute S = aQ as common secret.

The steps that Organization B performs are:
1. Choose a random integer b ∈ [1, r − 1].
2. Compute Q = bR in the group G, and send it to A.
3. Receive the element P ∈ G from A.
4. Compute S = bP as common secret.

Note that both A and B have computed the same values S, as
S = a(bR) = (ab)R = b(aR).

### B. Encryption/Decryption Scheme

For encryption and decryption we use ElGamal encryption scheme based on the discrete logarithm problem. We will describe it here, again for a general group G. [16]

Suppose that one has the following publicly known system parameters:
– The group G.
– An element R ∈ G of large prime order.

Organization A wants to send B a message M, which we assume to be encoded as an element of the group G. Organization A wants to encrypt M using B's public key Q, such that only B can decrypt the message again, using his secret key b.

To encrypt M, A does the following:
1. Obtain B's public key Q.
2. Choose a secret number a ∈ [1, r − 1].
3. Compute C1 = aR.
4. Compute C2 = M + aQ.
5. Send (C1, C2) to B.

B can decrypt the encrypted message by doing the following:
1. Obtain the encrypted message (C1, C2) from A.
2. Compute M = C2 − bC1.

### C. Signature Generation/Verification

The Digital Signature Algorithm can be used for any group G where the DLP is difficult, provided that one has a computable map G→Z with large enough image and few inverses for each element in the image. [16]

To create a key pair, A chooses a secret integer a ∈ Z, and computesP = aR. The number a is A's secret key, and P is her public key.

Assume the following system parameters are publicly known:
– A group G
– An element R ∈ G with large prime order r,
– A hash function H that maps messages m to 160-bit integers.

If A wants to sign a message m, it has to do the following:
1. Choose a random integer k ∈ [1, r − 1], and compute Q = kR.
2. Compute $s \equiv k^{-1}(H(m) + a\emptyset(Q)) \mod r$.

The signature is (m, Q, s).

To verify this signature, a verifier B has to do the following:
1. Compute $v_1 \equiv s^{-1}H(m) \mod r$ and $v_2 \equiv s^{-1}\emptyset(Q) \mod r$.
2. Compute $V = v_1R + v_2P$.
3. Accept the signature if V = Q. Otherwise, reject it.

## VIII. ADVANTAGES OF USING HCC IN CLOUDS

Hyperelliptic Curve Cryptosystem (HCC) is successor of (ECC) Elliptic Curve Cryptosystems. ECC proved to be better public key cryptosystems, because it creates smaller key sizes. The minimum key size for ECC should be 132 bits vs. 952 bits for RSA.

Hyperelliptic curves provide still smaller key size than ECC. So, smaller key sizes are having implementation advantages compared to RSA or ECC. Typical operand bit length for RSA is 1024…2048 bit, for ECC it is 160…256 bit, but for HCC it is 50…80 bits.

It is also perfectly suited for constraint environments like clouds, phones, smart cards. As a result it offers greater speed along with less storage.

As we know that end user has to pay for what he uses in clouds. That's make user to think about required resources i.e memory, processor etc. So what makes HCC in cloud desirable is that, HCC enabled devices will require less storage , less bandwidth.

## IX. CONCLUSION

As we know that cloud computing facing many security challenges. User is concerned about the data should remain confidential and also user needs authentication of other party while sending data over cloud. So we have presented schemes for key agreement, encryption/decryption and signature generation/verification using Hyperelliptic curve cryptography

Also HCC require less storage, less power and less bandwidth than other systems cryptosystems. Due to these properties HCC can be used in platforms that are constrained, such as thin clients. HCC creates smaller key sizes. It also helps in situations where efficiency is important.

. As for now HCC is still in development phase it may expand over genus ≥ 3.So in future these algorithms may reduce computation to further level.

## X. ACKNOWLEDGEMENT

The first author would like to thank Dr. Sushil Jajodia, Director, Center for Secure Information Systems, Volgenau School of Engineering, George Mason University, Fairfax, Virginia 22030-4422, USA for offering a Visiting Scholar position during Summer of 2014, when this work was carried on.